\documentstyle[twocolumn,aps,prl]{revtex}

\begin{document}
\draft
\title{ Reentrant Phase Diagram of Two-Dimensional Electron Gas}
\author{ Jukka A. Ketoja}
\address{ Department of Mathematics, P. O. Box 4,\\
FIN-00014 University of Helsinki, Finland}
\author{ Indubala I. Satija\cite{email}}
\address{
 Department of Physics and\\
 Institute of Computational Sciences and Informatics,\\
 George Mason University,\\
 Fairfax, VA 22030}
\date{\today}
\maketitle
\begin{abstract}
We study the phase diagram of an electron on a square lattice in
a transverse magnetic field.
Using a renormalization scheme, we show that the inequality of
the two next-nearest neighbor couplings destroys
the fat critical regime above the bicritical line
and replaces it with another reentrant extended phase.
Furthermore, the universal strange attractor describing self-similar
fluctuations of the localized phase
is replaced by a new fixed point. The reentrant extended phase and the
new fixed point of the localized phase belong to the universality
class of an anisotropic triangular lattice.

\end{abstract}

\narrowtext

As the properties of the Harper equation \cite{Harper},
 which describes an electron
on a square lattice in transverse magnetic field, are fairly
well understood \cite{Sok}, much less is known about more general quasiperiodic
tight binding models (TBM). In this paper, we study the
generalization which results from taking into account both the
nearest-neighbor (NN) and next-nearest-neighbor (NNN) interaction in
the electron problem.
The associated TBM has the form \cite{Thou}
\begin{eqnarray}
t_a (\psi_{k+1} +\psi_{k-1})
+2t_b \cos[2 \pi (k\sigma +\phi)] \psi_k\nonumber\\
+\exp[i2 \pi (k\sigma+\phi)] \{ t_{ab} e^{i\pi \sigma} \psi_{k+1} +
t_{a\bar b} e^{-i\pi \sigma} \psi_{k-1} \}
\nonumber\\
+\exp[-i2 \pi (k\sigma+\phi)] \{ t_{ab} e^{i\pi \sigma} \psi_{k-1} +
t_{a\bar b} e^{-i\pi \sigma} \psi_{k+1} \}\nonumber\\
 = E\psi_k
\end{eqnarray}
Here $t_a$ and $t_b$ are the NN couplings while $t_{ab}$ and $t_{a\bar b}$
are the diagonal NNN couplings.  The quasiperiodicity results from the
irrational parameter $\sigma$ describing the magnetic flux. The Harper
equation corresponds to vanishing of the NNN couplings.
Furthermore, in the limit $t_{a\bar b}$ equal to zero, the model

Recently, the above model was studied in the isotropic limit
where $t_{ab} = t_{a \bar b}$.\cite{Thou,KSC,KSloc}
 Although the wave
function $\psi_k$ in general is a complex function of the lattice index
$k$, $\psi_k$ can be taken to be real
in the case.
These studies revealed a very interesting phase
diagram :
For $2t_{ab} < t_a$, the model belonged to the universality class
of the Harper model with both extended (E) ($t_b < t_a$) and localized (L)
($t_b > t_a$) phases and a critical (C) point at $t_a = t_b$
where the system had the full square symmetry.
On the other hand, for $2t_{ab} \geq t_a$
the model was found to belong to a new
universality class where there was no E phase but
instead the C phase existed in a finite parameter interval
$t_b \leq 2t_{ab}$. For
$t_b > 2t_{ab}$ the states were exponentially localized.
The E and the C phases were separated by a bicritical line
$2t_{ab} =t_a$.

Detailed decimation studies \cite{KSC} showed that
the wave functions within the fat C phase above
the bicritical line were self-similar (at the band edges) only
at certain special values of the parameters.
These special points corresponded to
universal limit cycles of the renormalization. However, for generic
parameter values, the fractal characteristics of the critical wave
functions
did not exhibit self-similarity and were conjectured to be described
by a strange attractor of the renormalization flow.

In another interesting study of this model, it was shown that the
fluctuations of the wave functions in the L phase mimiced the behavior
in the C phase \cite{KSloc}. The L phase of the Harper universality
class was described by a renormalization
fixed point of the strong coupling limit
$t_b/t_a \to \infty$
while the L phase above the
bicritical line was described by a strange attractor of the
associated renormalization \cite{KSloc}.

The general case of the model where $t_{ab}$ and
$t_{a\bar b}$ are not equal and the resulting TBM is complex
has not been fully investigated. In the three parameter space,
$\lambda=t_b /t_a$, $\alpha=2t_{ab}/t_a$,
and $\beta= 2t_{a \bar b}/t_a$,  using the duality property of the model,
Han et al.\cite{Thou} calculated the Lyapunov exponent of the model
analytically and concluded
that the system is localized for $\lambda >1$ if
$(\alpha + \beta)/2 < 1$ and for $\lambda > (\alpha + \beta)/2$ otherwise.
Apart from the existence of metal-insulator transition, nothing is
known about the scaling
properties of the complex model.

In this paper, we study the complex model using our recently developed
decimation scheme.
We will confine ourselves to the case where $\sigma = (\sqrt 5 -1)/2$.
It is shown that the phase diagram changes
discontinuosly as $\alpha - \beta$ becomes different from zero.
The C phase of the real TBM above the bicritical line is replaced
by another E phase. This complex reentrant E phase is described
by a strong NNN coupling of the triangular lattice and is shown
to be related to the weak coupling limit of the Harper equation.
Furthermore, the invariant strange set of the renormalization describing
fractal characteristics of the fluctuations in the L phase degenerates
to a fixed point of the renormalization.
These fluctuations are defined by the equation \cite{KSloc}
\begin{equation}
\psi_k = e^{ -\gamma |k|} \eta_k
\end{equation}
where $\gamma$ is the Lyapunov exponent which vanishes in the E and
C phases and is positive in the L phase.
In other words, $\eta_k$ is equivalent to the
original wave function $\psi_k$ in the E and C phases whereas in
the L phase $\eta_k$ describes the fluctuations around the
exponentially decaying wave function. Knowing the analytic formula
for the Lyapunov exponent,
it is easy to write a TBM for $\eta_k$, resembling Eq. (1),
in the L phase \cite{KSloc}.

We will use a
decimation approach to describe the scaling properties of $\eta_k$,
the wave
function in the E and C phase and the fluctuations of the
wave functions in the L phase, for a specific value of energy.
In our studies below, we will focuss on
the quantum state with minimum energy $E_{min}$.
In addition to fixing the quantum state,
one has to also fix the phase factor $\phi$ to a critical
value in Eq. (1) so that the
wave function remains finite asymptotically.
The nondivergent wave functions are needed to determine the scaling properties
as has been discussed previously.
\cite{Ostlund,KS,KSC}

The key idea of the decimation scheme is to connect the wave function $\eta_k$
at
an arbitrary site $k$ with
two neighboring Fibonacci sites $k+F_{n+1}$ and $k+F_n$ where
$F_{n+1} = F_n + F_{n-1}$ \cite{KS,Ketoja}:
\begin{equation}
f_n(k) \eta(k+F_{n+1})=\eta(k+F_n) + e_n(k) \eta(k).
\end{equation}
The way how the decimation functions $f_n$ and $e_n$ are placed
in the decimation equation is somewhat arbitrary but here we choose
the form which causes the
asymptotic limits of the decimation functions $e_n$ and $f_n$ as
$n \to \infty$ to be bounded in all three phases.
The additive property of the Fibonacci numbers provides exact recursion
relations for the decimation functions $e_n$ and $f_n$ \cite{KSloc,KS,Ketoja}:
\begin{eqnarray}
e_{n+1} (k)&=& - {A e_n (k) \over 1+Af_n (k)} \\
f_{n+1} (k)&=& {f_{n-1} (k+F_n) f_n(k+F_n)\over 1+Af_n(k)} \\
A &=& e_{n-1} (k+F_n) + f_{n-1} (k+F_n)e_n(k+F_n). \nonumber
\end{eqnarray}
For fixed $k$, the above coupled equations for the decimation
functions define a RG flow which asymptotically ($n \rightarrow \infty$)
converges on an attractor. The C phase is distinguished from E phase
by the existence of  nontrivial limiting behavior. With anisotropic NNN
coupling, the attractor
is a $p$-cycle in all three phases for $E=E_{min}$ \cite{footnote0}.
The asymptotic
$p$-cycle for $e_n (0)$ and $f_n (0)$ determines the universal scaling ratios
\begin{equation}
\zeta_j = \lim_{n \rightarrow \infty} \eta(F_{pn+j})/\eta(0);\;\;j=0,...,p-1.
\end{equation}
whose absolute values
 are equal to unity in the E phase and less than unity in the C
phase \cite{footnote0}.

In Fig. 1 we show the phase diagram obtained by analyzing the asymptotic
behavior of the decimation functions in different parts of the parameter
space. As soon as $\alpha$ and $\beta$ differ, $\alpha-\beta$
is an irrelevant parameter and the phase diagram is determined solely
by two parameters: $\lambda$ and $(\alpha+\beta)/2$.
In the parameter range $(\alpha +\beta)/2 < 1$, the decimation
functions
become asymptotically real approaching the same universal cycles as
for the Harper equation $\alpha =\beta = 0$ \cite{KS,KSloc}. For
$(\alpha +\beta)/2 \geq 1$, $\alpha \not= \beta$,
 the decimation functions stay complex also
asymptotically. The cycle length $p$ in this case is six in all three
phases. However, considering the absolute value of the decimation
functions (or the scaling ratio $\zeta$), one observes a 3-cycle
on the line $(\alpha +\beta)/2 =1$ and a fixed point for
$(\alpha + \beta)/2 > 1$. The 3-cycle observed at the critical point C is
different from the universal 3-cycle observed along the rest of the
(bi)critical line AC. Moreover, these fixed points
are all universal and do not depend
on the actual parameter values. Table I summarizes various universality
classes of the model. It is interesting to note that the line
CL divides the localized phase into two different universality classes.
That is, the scaling properties of the self-similar fluctuations in the
region BCL are different from those of the region LCE. Furthermore, the
boundary line CL has its scaling characteristics different from the
two regions that it separates.

The reentrant E phase described by a complex $6$-cycle with
 complex scaling ratios
is not as trivial as the real fixed point of the E
phase below the bicritical line describing the weak coupling limit
of the Harper model.
The latter depends neither on the phase $\phi$
nor on the lattice index $k$ and can be easily solved from a
fixed point equation: $f_n (k) \equiv \sigma$ and $e_n (k) \equiv
-\sigma^2$ resulting in $\zeta =1$. In the complex E phase, complications
arise from the fact that the
decimation functions do depend both on $\phi$ and $k$.
But the absolute value of a decimation function has the same
constant value as in the real case.
Noting the fact that the decimation
functions and the scaling ratios above the line ACL
 remain the same, the reentrant E phase can be understood in
the limit of highly anisotropic triangular lattice
$\alpha \rightarrow \infty$, $\beta/\alpha \rightarrow 0$, and
$\lambda/\alpha \rightarrow 0$. In this limit, Eq. (1)
reduces to the TBM describing the weak coupling
limit of the Harper model,
\begin{equation}
C_{k+1} + C_{k-1} = {E\over t_{ab}} C_k,
\end{equation}
where $C_k$ is related to $\psi_k$ via
\begin{equation}
C_k = \exp(i2\pi\phi k)\exp(i\pi \sigma k^2) \psi_k .
\end{equation}
Hence, the wave function of the infinitely anisotropic triangular lattice
is related by a phase
factor to the extended wave function of the similar NN square lattice.
The above equation shows that the scaling factors $\zeta_j$ exist only
for special (i.e. rational or those related to the golden mean)
 values of the phase $\phi$.
The relation $F_n \sigma = F_{n-1} - (-\sigma)^n$ implies
that
\begin{equation}
\sigma F_n^2 = F_{n-1} F_n + {(-1)^{n-1} + \sigma^{2n} \over
1+2\sigma}.
\end{equation}
 From this and Eq. (8) it follows that for $\phi =0,1/2$
\begin{equation}
%\zeta_j = \pm \cos[\pi /(1+2\sigma)] \pm i\sin[\pi /(1+2\sigma)] .
\zeta_j = \pm \exp[ \pm i\pi/(1+2\sigma) ]
\end{equation}

The above scaling is relevant also in other
parts of the complex E phase because the RG flow is attracted by
the same $6$-cycle in the whole region.
Taking advantage of the above limiting solution,
it is possible to derive an explicit expression for the $6$-cycle.
Substituting Eq. (8) into the decimation equation (3),
we obtain
\begin{eqnarray}
e_n (k)&=& e_n^h (k) \exp(-i 2\pi \phi F_n) \exp[-i\pi \sigma (F_n^2 +2kF_n)]
\nonumber\\
f_n (k)&=& f_n^h (k) \exp(i2\pi \phi F_{n-1}) \nonumber\\
&\times& \exp[i\pi \sigma (F_{n+1}^2 -F_n^2 +2kF_{n-1})],
\end{eqnarray}
where $e_n^h$ and $f_n^h$ are the decimation functions corresponding
to the TBM (7) describing the weak coupling limit of the Harper equation.
 From these equations we see that $e_n$ and $f_n$
are functions of the fractional part of $k \sigma $, denoted by
$<k \sigma >$, only. Therefore, we can write the decimation functions
 in terms of the
renormalized variable $x=(-\sigma)^{-n} <k \sigma >$ \cite{KS}. For
simplicity, let us assume that $\phi =0,1/2$.
Applying the relation (9) and the fact that
\begin{equation}
(-\sigma)^n F_n = {(-1)^n - \sigma^{2n} \over 1+2\sigma}
\end{equation}
we obtain six different limiting function pairs of the form
\begin{eqnarray}
e^* (x) = \pm \sigma^2 \exp[\pm i\pi (2x-1)/(1+2\sigma)] \nonumber\\
f^* (x) = \pm \sigma \exp[\pm i\pi 2(\sigma x+1)/(1+2\sigma)]
\end{eqnarray}
as $n$ tends to infinity. These pairs form a 6-cycle of
the recursion (4-5), written in the continuos variable $x$ \cite{KS}:
\begin{eqnarray}
e_{n+1} (x)= - {A e_n (-\sigma x) \over 1+Af_n (-\sigma x)} \\
f_{n+1} (x)= {f_{n-1} (\sigma^2 x+\sigma) f_n(-\sigma x-1)\over
1+Af_n(-\sigma x)} \\
A = e_{n-1} (\sigma^2 x+\sigma) + f_{n-1} (\sigma^2
x+\sigma)e_n(-\sigma x -1). \nonumber
\end{eqnarray}
For the above 6-cycle,
$1+Af^*(-\sigma x) \equiv \sigma$.

Therefore, the characteristic feature of the reentrant E phase
is the fact that the decimation functions depend explicitly on $x$.
These functions are complex and consequently the universal scaling
ratio has both real and imaginary parts but the absolute value of
$\zeta$ is unity.
This is unlike the Harper E phase where the real universal functions
are site independent and are given in terms of the powers of the
 golden mean.

In summary, we have shown that
with anisotropic NNN couplings, the renormalization behavior for the
TBM describing an electron on a square lattice
is a lot simpler than in the isotropic case. Firstly, the renormalization
strange set corresponding to the fat C phase in the isotropic case
is replaced by an attracting cycle associated with trivial scaling
properties (i.e. E phase). Secondly, the bicritical lines of the isotropic
case remain critical also when the NNN couplings are not equal
but the renormalization attractor is again simpler (i.e. a cycle).
Thirdly, the fluctuations of the
exponentially localized wave functions are described by a
universal fixed point and not by an infinite strange set as in
the isotropic case.

The novel feature of our model is the existence of two extended
phases separated by a critical line signaling a transition from
 real scaling ratios to complex ones. The two extended
phases respectively fall into the universality classes of
 the weak coupling limit of the NN square lattice
and the strong NNN coupling limit of the triangular lattice.
It is interesting that also on the other side of the
localization boarder (i.e. in the L phase),
 the scaling behavior is divided into three different universality classes.
The Harper L phase corresponding to the strong NN coupling
is separated from the strong NNN coupling phase by the CL line with its
own scaling properties. Therefore, for a fixed value of NNN coupling,
as the NN coupling $\lambda$ is varied,
the behavior beyond the localization transition
appears to be shadowed by the behavior before the localization transition:
the existence of E phase before localization results in a nontrivial
fixed point in the L phase while the existence of a period $p$
limit cycle describing self-similar critical states before the localization
onset
leads to the appearence of period $p$ limit cycle of the
renormalization after localization describing self-similar fluctuations.

The phase diagram discussed here describes the universal properties
of square and triangular lattices . Therefore,
we believe that our results can be
experimently realized on two-dimensional mescoscopic systems.\cite{expt}

\acknowledgements

We are grateful to Siamak Khodaei for computing the energies of the complex
 model
to high precision and verifying the phase diagram using total bandwidth
scaling criterion.
The research of IIS is supported by a grant from National Science
Foundation DMR~093296. JAK would like to thank the Niilo Helander
Foundation for financial support. IIS would like to acknowledge the
hospitality of Magnetic Materials group at National Institute of
Science and Technology where part of this research is done.

\begin{figure}
\caption{ The reentrant phase diagram of the TBM (1) for the case
$\alpha \not= \beta$.
The (bi)critical line AC separates the two extended phases while the
critical line
CE describes the onset to localization. The extension of the
AC line (line CL) also separates the localized phase
into two different regions with different scaling properties.
 For the corresponding phase
diagram in the isotropic case $\alpha=\beta$, see Fig. 1 of ref.
[5].}
\label{fig1}
\end{figure}

\begin{table}
\caption{The universal scaling ratios for various universality classes
at the band edge. In order to avoid showing all $p$ scaling ratios,
some of which differ only by signs of the real or imaginary part,
we show all different $(|Re(\zeta_j)|,|Im(\zeta_j)|)$.
$\phi_c =1/2$ for all other parts of the parameter space
except the line AC where $\phi_c$ varies as a function of
$\lambda$.
In the reentrant E region ACE, the absolute value of $\zeta$
is unity. Note that the last three lines show the scaling properties
of the fluctuations in an exponentially localized wave function.}
\begin{tabular}{ccc}
 & & \\
Region & $\zeta$ & Nature of Phase\\
 & & \\
\tableline
& & \\
Region BCA & (1, 0) & Real E\\
 & & \\
Line BC & (.211, 0) & Real C\\
& & \\
Point C & (.181, .154)& Complex C\\
 &(.196, .231)& \\
 &(.222, .188)& \\
& & \\
Line AC &(.587, .694) & Complex C\\
 &(.638, .540) & \\
 &(.630, .533) & \\
& & \\
Line CE & (.0348, .208) & Complex C\\
& & \\
Region ACE & (.165, .986) & Complex E\\
& & \\
Region BCL & (.176, 0) & Real L\\
& & \\
Line CL & (.173, .204) & Complex L\\
 &(.237, .200)& \\
 &(.0928, .0785) & \\
& & \\
Region LCE & (.0285, .170) & Complex L\\
& & \\
\end{tabular}
\end{table}


\begin{references}
\bibitem{footnote0} The $p$-cycle is observed for the phase $1/2$ (or $0$)
and for the critical phase $\phi_c$ for which
the main peak of the wave function is at $k=0$ (in many cases
$\phi_c =1/2$).
However, for an arbitrary phase the asymptotic behavior of the
decimation
functions is more complicated. In the numerical work, we have used
this cycle property to improve the initial estimate for $E_{min}$
obtained by diagonalizing a large matrix (see ref. \cite{KS}).
\end{references}
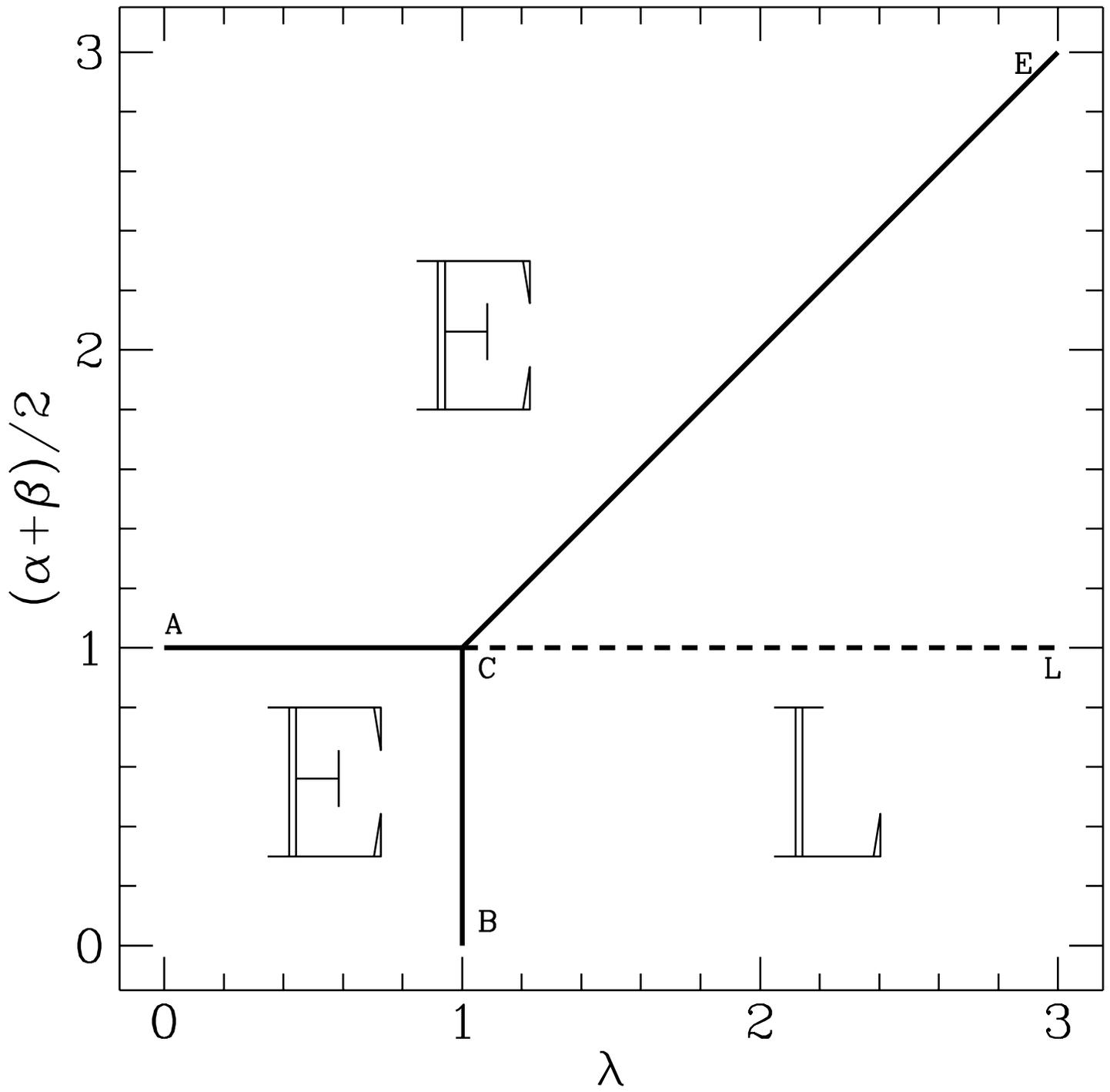
\end{document}